\documentclass[
 reprint,
 groupedaddress,
 amsmath,amssymb,
 aps,
 prb,
]{revtex4-1}

\usepackage{float}
\usepackage{graphicx}
\usepackage{dcolumn}
\usepackage{bm}
\usepackage{xcolor}
\usepackage[normalem]{ulem}
\usepackage[colorlinks=true,allcolors=blue]{hyperref}
\usepackage{braket}

\newcommand{\omegage}{\omega_\text{ge}}
\newcommand{\omegamone}{\omega_{\text{m}_1}}
\newcommand{\omegamtwo}{\omega_{\text{m}_2}}
\newcommand{\omegami}{\omega_{\text{m}_i}}

\newcommand{\Tonem}{T_{1,\text{m}}}
\newcommand{\Ttwom}{T_{2,\text{m}}}
\newcommand{\Tonemone}{T_{1,\text{m}_1}}
\newcommand{\Tonemtwo}{T_{1,\text{m}_2}}
\newcommand{\Ttwomone}{T_{2,\text{m}_1}}
\newcommand{\Ttwomtwo}{T_{2,\text{m}_2}}

\newcommand{\FidOne}{\mathcal{F}_s}
\newcommand{\FidTwo}{\mathcal{F}_\text{Bell}}

\newcommand{\Xpi}{\hat{X}_\pi}
\newcommand{\Xpihalf}{\hat{X}_{\pi/2}}

\newcommand{\Ypihalf}{\hat{Y}_{\pi/2}}

\newcommand{\ketBell}{\ket{\psi_{\text{Bell}}}}
\newcommand{\braBell}{\bra{\psi_{\text{Bell}}}}

\newcommand\Tstrut{\rule{0pt}{3.25ex}}
\newcommand\Bstrut{\rule[-1.25ex]{0pt}{0pt}}

\begin{document}

\title{Quantum state preparation, tomography, and entanglement of mechanical oscillators}

\author{E. Alex Wollack}
\thanks{These authors contributed equally to this work}
\author{Agnetta Y. Cleland}
\thanks{These authors contributed equally to this work}
\author{Rachel G. Gruenke}
\author{Zhaoyou Wang}
\author{Patricio Arrangoiz-Arriola}
\author{Amir H. Safavi-Naeini}
\thanks{\href{mailto:safavi@stanford.edu}{safavi@stanford.edu}}
\affiliation{Department of Applied Physics and Ginzton Laboratory, Stanford University\\348 Via Pueblo Mall, Stanford, California 94305, USA}

\date{\today}

\maketitle

\textbf{Precisely engineered mechanical oscillators keep time, filter signals, and sense motion, making them an indispensable part of today's technological landscape. These unique capabilities motivate bringing mechanical devices into the quantum domain by interfacing them with engineered quantum circuits. Proposals to combine microwave-frequency mechanical resonators with superconducting devices suggest the possibility of powerful quantum acoustic processors~\cite{Pechal2018,Hann2019,Chamberland2020}. Meanwhile, experiments in several mechanical systems have demonstrated quantum state control and readout~\cite{Satzinger2018,Chu2018}, phonon number resolution~\cite{Arrangoiz-Arriola2019,Sletten2019}, and phonon-mediated qubit-qubit interactions~\cite{Bienfait2019,Bienfait2020}. Currently, these acoustic platforms lack processors capable of controlling multiple mechanical oscillators' quantum states with a single qubit, and the rapid quantum non-demolition measurements of mechanical states needed for error correction. Here we use a superconducting qubit to control and read out the quantum state of a pair of nanomechanical resonators. Our device is capable of fast qubit-mechanics swap operations, which we use to deterministically manipulate the mechanical states. By placing the qubit into the strong dispersive regime with both mechanical resonators simultaneously, we determine the resonators' phonon number distributions via Ramsey measurements. Finally, we present quantum tomography of the prepared nonclassical and entangled mechanical states. Our result represents a concrete step toward feedback-based operation of a quantum acoustic processor.}

The burgeoning field of quantum acoustics combines the established tools and infrastructure of circuit quantum electrodynamics (cQED) with the many benefits of nanomechanical oscillators. This creates a rich platform for explorations of fundamental quantum physics \cite{OConnell2010,Arrangoiz-Arriola2016,Chu2017,Chu2018,Satzinger2018,Arrangoiz-Arriola2019,Sletten2019}, with promising applications toward scalable quantum computation~\cite{Chu2020,Pechal2018,Hann2019}. Over a small footprint, mechanical systems have the potential to provide access to a large number of highly coherent microwave-frequency modes which can act as high-precision sensors of force and motion, store long-lived quantum memories with minimal crosstalk, and form interconnects with optical systems. Furthermore, it is possible to generate nonclassical~\cite{Satzinger2018, Chu2018} and entangled states of motion~\cite{Jost2009,Ockeloen-Korppi2018,Riedinger2018,Barzanjeh2019,Mercier2021,Kotler2021} in mechanical oscillators, making them a compelling system for storing and processing quantum information. By placing these acoustic systems in the strong dispersive coupling limit~\cite{Bertet2002,Schuster2007}, both non-Gaussian and non-demolition measurements can be made via phonon-number resolved detection.

Access to this regime is enabled by our device design and heterogeneously integrated material platform.  We leverage the small mode volume and strong piezoelectricity of a phononic crystal resonator in thin-film lithium niobate (LN), combined with a high coherence aluminum transmon qubit, to achieve large coupling rates between a superconducting qubit processor and two nanomechanical resonators. Our approach allows for strong coupling while suppressing the phonon radiation loss channels that arise in piezoelectric materials. In phononic crystal devices, the density of states for acoustic radiation loss can be eliminated over a wide frequency range by choosing a periodic geometry that produces a full phononic bandgap~\cite{Arrangoiz-Arriola2016}. This approach localizes the gigahertz frequency mechanical mode to a wavelength-scale volume~\cite{ArrangoizArriola2018}, and has produced resonators with extremely long mechanical lifetimes~\cite{MacCabe2019}. With improved fabrication processes (see methods), we have extended both the qubit and mechanical resonators' coherence times, $T_1$ and $\Tonem$, which limited experimental capabilities in prior work~\cite{Arrangoiz-Arriola2019}.

\begin{figure}[!htb]
    \centering
    \includegraphics[width=89mm]{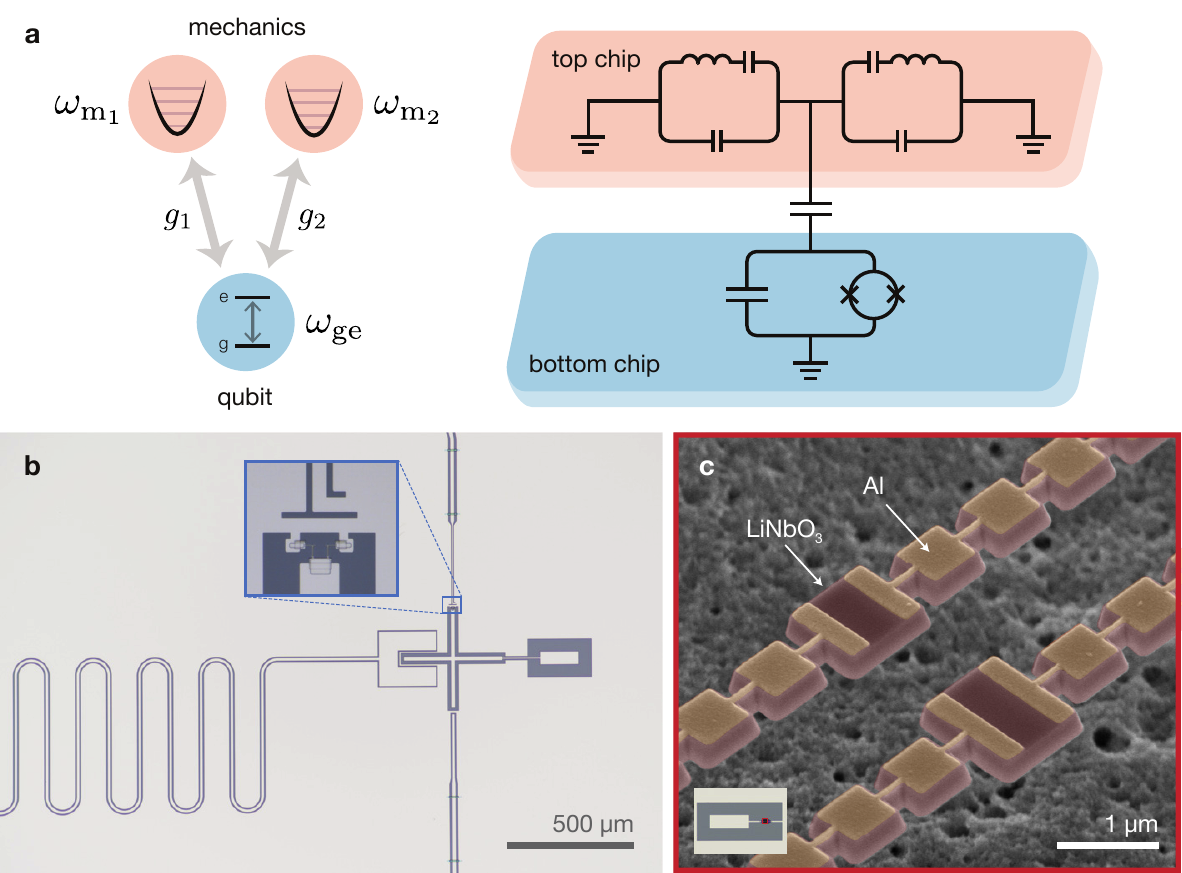}
    \caption{{\bf Device description.} 
    {\bf a,}~Schematic of the modes and flip-chip device. A frequency-tunable qubit on the bottom chip (blue) is capacitively coupled through a small vacuum gap to two mechanical modes on the top chip (orange). The mechanical modes are represented as Butterworth-van Dyke equivalent circuits. 
    {\bf b,}~Optical micrograph of the bottom (qubit) chip, with inset showing the qubit's SQUID and adjacent flux-line, used for frequency control. The rightmost arm of the transmon island extends to form the bottom pad of the coupling capacitor. 
    {\bf c,}~False-color scanning electron micrograph of the top (mechanics) chip, showing two phononic crystal resonators (red). Aluminum electrodes (orange) are galvanically connected both to the top chip’s coupling capacitor pad and ground plane, as shown in the inset.}
    \label{fig_device}
\end{figure}

Our hybrid device is composed of two chips integrated in a flip-chip architecture~\cite{Satzinger2019} (Fig.\,\ref{fig_device}a). We fabricate a frequency-tunable transmon qubit~\cite{Kelly2015} with microwave control lines and a coplanar waveguide readout resonator (Fig.\,\ref{fig_device}b) on a $6\,\textrm{mm}\times 9\,\textrm{mm}$ silicon chip. The qubit is capacitively coupled through a small vacuum gap to two phononic crystal resonators fabricated on a separate $2\,\textrm{mm}\times 4\,\textrm{mm}$ top chip (Fig.\,\ref{fig_device}c). These cavities are patterned by argon ion milling a thin film of LN~\cite{Wang2014}, which is then released from the chip’s silicon handle by a xenon difluoride dry etch~\cite{Arrangoiz-Arriola2019, Vidal-alvarez2017}. Each mechanical eigenmode is confined to a small defect site suspended on either side by a one-dimensional phononic crystal mirror~\cite{Arrangoiz-Arriola2016, ArrangoizArriola2018}. Utilizing the piezoelectric effect of LN, the qubit couples to the mechanical modes via aluminum electrodes patterned on each resonator. These electrodes extend to a metallized pad which forms the top half of a cross-chip coupling capacitor, with a matching pad on the qubit island. The capacitor gap is defined by the flip-chip separation distance of 1\,$\mu$m (see methods for flip-chip procedure). 

The Hamiltonian for the resulting device includes two mechanical oscillators with frequencies $\omegami$ and lowering operators $\hat{b}_i$, in addition to a qubit with transition frequency $\omegage$ and Pauli operators $\hat{\sigma}$: $\hat{H}_0 =  \omegamone\hat{b}_1^\dagger \hat{b}_1 + \omegamtwo\hat{b}_2^\dagger  \hat{b}_2 +\frac{1}{2}\omegage\hat{\sigma}_z$. A direct piezoelectric coupling between the qubit and mechanics leads to an interaction Hamiltonian $\hat{H}_\text{int} = \sum_i g_i  (\hat{b}_i+\hat{b}^\dagger_i)\hat{\sigma}_x$,  with coupling rates $g_i$. In the limit of large detuning, the interaction is best described by an effective dispersive Hamiltonian~\cite{Koch2007} 
\begin{eqnarray*}
\hat{H}_{\text{eff}} &=& \hat{H}_0 + (\chi_1 \hat{b}_1^\dagger \hat{b}_1 + \chi_2 \hat{b}_2^\dagger \hat{b}_2)\hat{\sigma}_\text{z}\,.
\end{eqnarray*}
In this regime, each mechanical mode imparts a frequency shift of $2 \chi_i$ per phonon on the qubit. This dispersive coupling rate $\chi_i$ is related to the qubit anharmonicity $\alpha_q$, each mechanical mode's coupling rate $g_i$, and the detunings $\Delta_i = \omegage - \omegami$ by~\cite{Koch2007} 
$$\chi_i = -\frac{g_i^2}{\Delta_i} \frac{\alpha_q}{\Delta_i - \alpha_q}\,.$$

The time required to resolve these phonon-induced frequency shifts is roughly $\pi/\chi_i$, making it important for $\chi_i$ to exceed the decoherence rates of both the mechanical resonators and the qubit. A system that satisfies this condition, while maintaining the detuning requirement $\Delta_i\gg g_i$ for the effective Hamiltonian to hold, is said to be in the strong dispersive coupling regime~\cite{Schuster2007}, which has only recently been demonstrated for circuit quantum acoustic devices~\cite{Sletten2019,Arrangoiz-Arriola2019}. A useful figure of merit for devices in this regime is the dispersive cooperativity $C = 4\chi^2 T_1 \Tonem$, which our device improves to $C=490$ compared with $C = 170$ in previous work in quantum acoustics~\cite{Arrangoiz-Arriola2019}.

\begin{figure}[!htb]
    \centering
    \includegraphics[width=89mm]{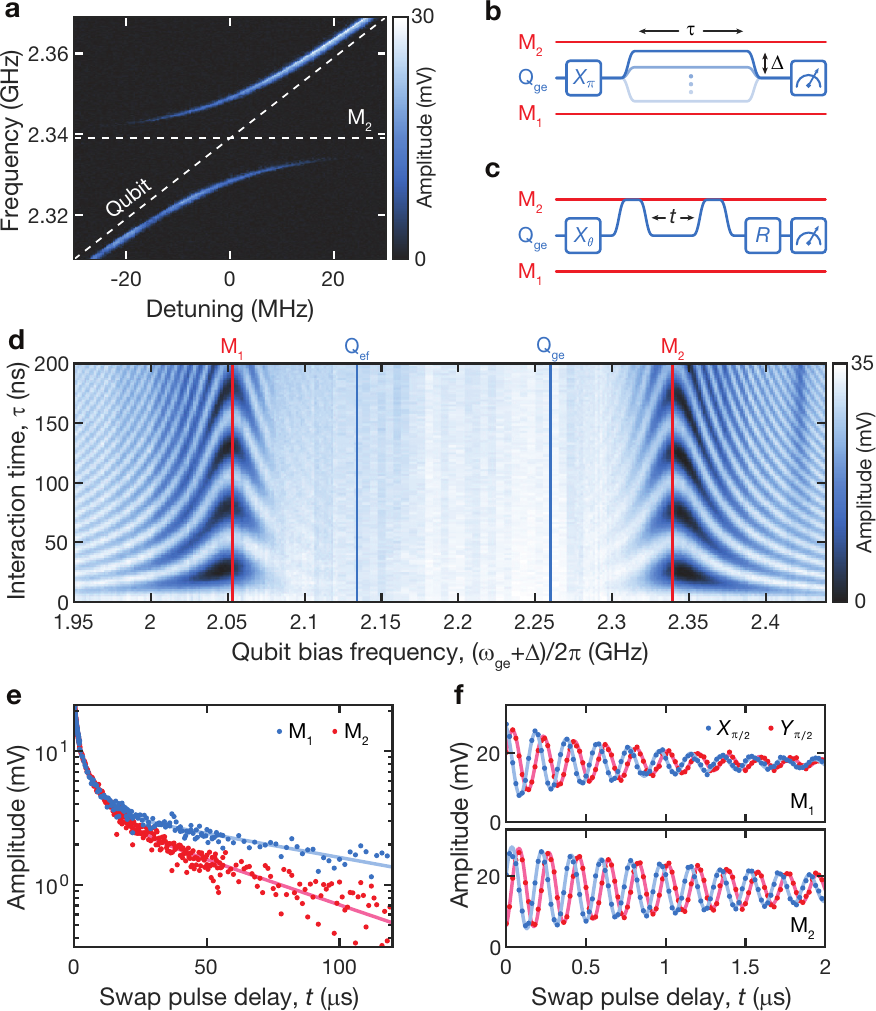}
    \caption{{\bf Characterization of the mechanical modes.}  
    {\bf a,}~Qubit spectroscopy near the mechanical mode $M_2$, with $\omegage$ detuned relative to $\omegamtwo$.
    {\bf b,}~Pulse sequence for Rabi-swap experiment. The qubit is excited to $\ket{e}$ using a $\Xpi$ pulse, then flux-detuned by frequency $\Delta$ for an interaction time $\tau$ before qubit measurement.
    {\bf c,}~Pulse sequence for single-phonon $\Tonem$ and $\Ttwom$ experiments. The qubit is prepared using either a $\Xpi$ or $\Xpihalf$ rotation, then swapped to one of mechanical modes $M_1$ or $M_2$. After waiting a variable delay time $t$, the qubit and mechanics are swapped again, followed by an optional qubit tomography rotation $R$ and measurement.
    {\bf d,}~Qubit response as a function of bias frequency $\omegage + \Delta$ and interaction time $\tau$ of the applied flux pulse in {\bf b}. At the start of the experiment, the qubit is held at $\omegage/2\pi = 2.26\,\text{GHz}$ (rightmost blue line, where $\Delta = 0$) before being frequency-detuned to interact with the mechanical modes (red lines).
    {\bf e,}~Single phonon $\Tonem$ measurement for each mechanical mode (blue: $M_1$, red: $M_2$), using the pulse sequence in {\bf c} with the qubit prepared in $\ket{e}$ and the identity operation for $R$.
    {\bf f,}~Single phonon $\Ttwom$ measurements of the mechanical modes (top: $M_1$, bottom: $M_2$). Here, the qubit is initially prepared in the superposition $\ket{g} + \ket{e}$, and we use tomography rotations $R = \Xpihalf$ (blue) or $\Ypihalf$ (red) in {\bf c}.}
    \label{fig_characterization}
\end{figure}

For this experiment, we leverage established techniques in cQED to perform state preparation and readout of the qubit (see methods), allowing characterization of the mechanical resonators using the qubit as a probe. We control the qubit frequency by flowing current through an on-chip flux-line shown in Fig\,\ref{fig_device}b. Tuning the qubit yields avoided crossings in the qubit spectrum at both the lower and upper mechanical frequencies, $\omegamone/2\pi = 2.053\,\text{GHz}$ and $\omegamtwo/2\pi = 2.339\,\text{GHz}$ (Fig.\,\ref{fig_characterization}a). From these avoided crossings, we determine the qubit-mechanics coupling strengths to be $g_1/2\pi=(9.5{\,\pm\,}0.1)\,\text{MHz}$ and $g_2/2\pi=(10.5{\,\pm\,}0.1)\,\text{MHz}$. 

Although the static capacitive coupling between the qubit and mechanics is fixed, the qubit-mechanics interaction is controlled on nanosecond timescales by rapidly tuning the frequency of the qubit between the off-resonant ($|\omegage-\omegami|\gg g_i$) and on-resonant ($\omegage=\omegami$) regimes via current pulses sent through the flux-line. To characterize and calibrate swap operations, we bias the qubit frequency to $\omegage/2\pi=2.26\,\text{GHz}$, far from the mechanical resonances. Using the pulse sequence of Fig.\,\ref{fig_characterization}b, we perform Rabi-swap experiments using a single initial excitation in Fig.\,\ref{fig_characterization}d. At the correct detunings, the excitation is exchanged between the qubit and one mechanical resonator, enabling transfer of the qubit state to the mechanics. We perform an $i\textsc{swap}$ operation in a time of $\pi/2g_i \simeq 24-26\,\text{ns}$, and estimate a fidelity of $0.95{\,\pm\,}0.01$ from the fringe visibility.

Access to a fast, high-fidelity swap operation allows us to extend our control of the qubit to the mechanical devices. We perform single-phonon characterization of both resonators using the pulse sequence in Fig.\,\ref{fig_characterization}c. In these experiments, we use the qubit to prepare a quantum state of the resonator, then wait a delay time $t$ before swapping the mechanical state back into the qubit for measurement. By choosing to initially rotate the qubit into the state $\ket{e}$ or $\ket{g}+\ket{e}$, we  characterize either the mechanical energy decay time $\Tonem$ or mechanical dephasing time $\Ttwom$.

We observe that both resonators exhibit energy relaxation dynamics that are best described as the sum of three decaying exponentials (Fig.\,\ref{fig_characterization}e). The fastest decay is observed to be $\Tonemone = (1.23{\,\pm\,}0.08)\,\mu\text{s}$ and $\Tonemtwo = (0.99{\,\pm\,}0.03)\,\mu\text{s}$ for mechanical resonators $M_1$ and $M_2$. In contrast, the other decay times are on the order of $10$ and $90\,\mu\text{s}$ for both resonators. The observed multi-exponential response may be explained by resonant decay into saturable and rapidly dephasing two-level systems (TLS) in the device~\cite{Wollack2021,Heidler2021}, but a more detailed study is required.

The results of a similar Ramsey experiment are shown in Fig.\,\ref{fig_characterization}f and used to extract the mechanical dephasing times $\Ttwomone = (0.87{\,\pm\,}0.02)\,\mu\text{s}$ and $\Ttwomtwo = (1.71{\,\pm\,}0.03)\,\mu\text{s}$. For a harmonic oscillator under the presence of amplitude damping, we expect each mechanical resonator's $\Ttwom$ to be twice its $\Tonem$; however, both modes seem to suffer from an additional, non-negligible source of phase decoherence, with inferred pure dephasing times $T_{\phi,\text{m}_1}=1.4\,\mu\text{s}$ and $T_{\phi,\text{m}_2}=13\,\mu\text{s}$. This may be also due to the presence of TLS, and a more complete analysis of decoherence in these devices will be the subject of future studies.

After characterizing the device, we use the qubit to perform full quantum state tomography of the upper mechanical resonator. Our goal is to obtain the density matrix $\hat{\rho}$ describing a single resonator's state. Previously, this has been achieved through dynamics where the qubit and mechanics directly exchange excitations~\cite{Satzinger2018, Chu2018}. Here, we use the strong dispersive interaction to impart a phonon-number dependent frequency shift on the qubit, which is then read out by a Ramsey measurement~\cite{Lachance2020, Gambetta2006,Schuster2007,Brune1990,Brune1994,Bertet2002} that yields the phonon number distribution $P_0(n)$. This provides us with the diagonal elements of the density matrix $\langle n |\hat{\rho}|n\rangle$, but does not fully determine the state. To gain information about $\hat \rho$'s off-diagonal elements, we perform a calibrated displacement operation $\hat D_\alpha$ on the mechanical resonator before the Ramsey measurement to find $P_\alpha(n)\equiv \langle n| \hat D_\alpha \hat \rho \hat D^\dagger_\alpha|n\rangle$.

\begin{figure*}[!htb]
    \centering
    \includegraphics[width=183mm]{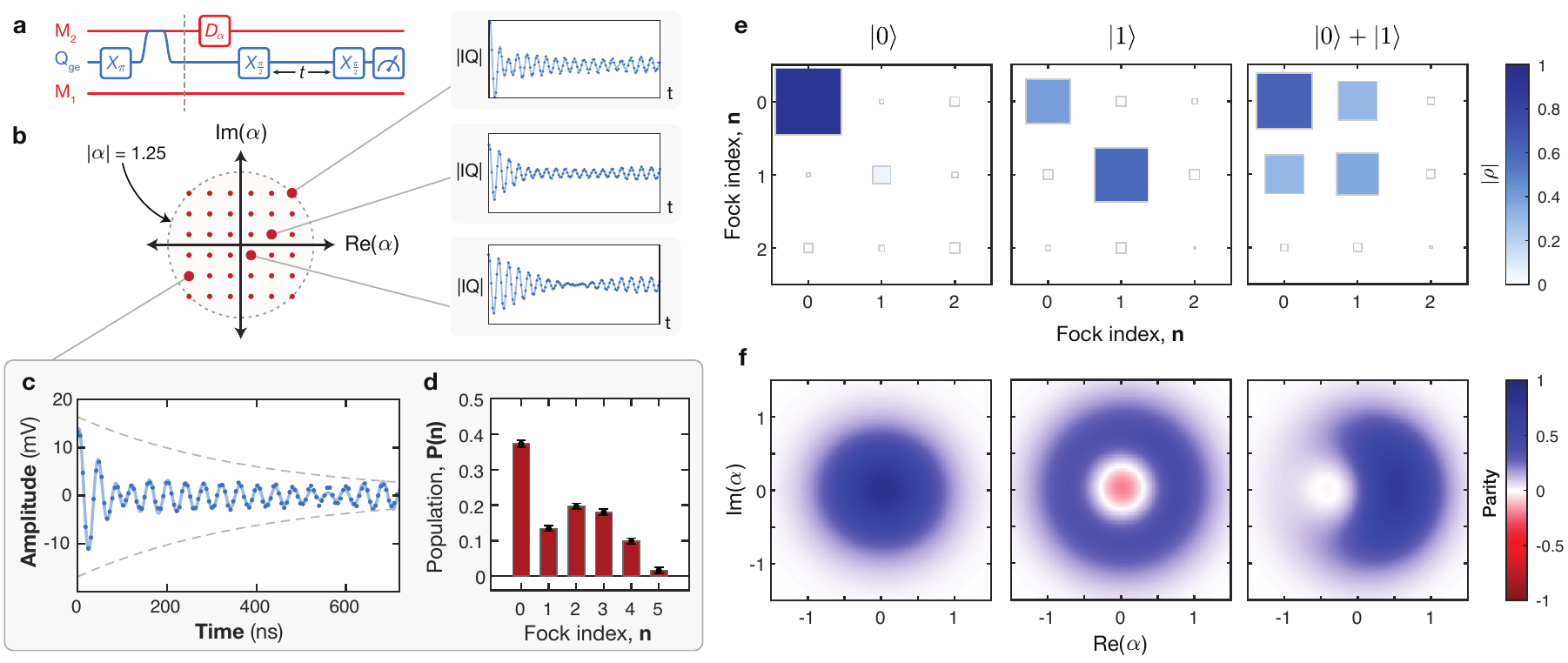}
    \caption{{\bf Single-mode tomography.} {\bf a,}~Pulse sequence showing state preparation, displacement, and Ramsey measurement. First, we use the qubit (blue) to prepare $\ket{1}$ in the upper mechanical mode, $M_2$. $M_2$ is then displaced by a microwave pulse $\hat{D}_\alpha$ with variable amplitude and phase. Finally, we perform a Ramsey sequence on the qubit. For $\ket{0}$, state preparation (left) is omitted, and for $\ket{0} + \ket{1}$ the $\Xpi$ pulse is replaced with $\Xpihalf$.
    {\bf b,}~Complex-valued amplitudes $\alpha$ of the displacements $\hat{D}_\alpha$ (red points), with a few corresponding measurement results (highlighted points).
    {\bf c,}~Representative Ramsey measurement result and {\bf d,}~extracted phonon number distribution. The data (dark blue points) are fit to Eq.\,\ref{eq_ramsey_fit} (light blue line) with the grey dashes showing the fitted decay envelope.
    {\bf e,}~Reconstructed density matrices and {\bf f,}~Wigner functions for each prepared state, extracted by convex optimization.}
    \label{fig_one_mode_tomo}
\end{figure*}

We begin the tomography protocol by using the qubit to prepare phonon states $\ket{0}$, $\ket{1}$, or $\ket{0} + \ket{1}$ in the upper mechanical mode. For this experiment, the qubit is initially biased to $\omegage/2\pi = 2.26\,\text{GHz}$ to ensure sufficient detuning for a dispersive interaction, $|\Delta_2|/g_2 \simeq 8$. We synthesize these states by first rotating the qubit to the desired state with an $\Xpi$ or $\Xpihalf$ pulse, then swapping it into the resonator, as shown in Fig.\,\ref{fig_one_mode_tomo}a. Next, we displace the resonator state ($\hat D_\alpha$) with a microwave pulse  at the mechanical frequency, applied to the qubit's $XY$ line~\cite{Arrangoiz-Arriola2019}. We then perform a Ramsey measurement to resolve the dispersive shifts on the qubit resulting from each populated Fock level in the displaced mechanical state. The resulting signal takes the form of a sum of oscillating terms with an exponentially decaying envelope,
\begin{align} \label{eq_ramsey_fit}
    S(t) = \sum_{n=0} A_n e^{-\kappa t/2} \cos [(\omega_0 + 2 \chi n) t + \varphi_n]\,.
\end{align}
We fit the data to Eq.\,\ref{eq_ramsey_fit} to learn the weight $A_n$ and frequency of each spectral component (see methods) and measure a dispersive shift $\chi/2\pi=(\text{-}718{\,\pm\,}7)\,\text{kHz}$. This fit allows us to extract the population $P_\alpha(n)$ in each Fock level $n$ by normalizing the spectral amplitudes: $P_\alpha(n) = A_n / \Sigma_n A_n$. A representative time-domain fit and extracted distribution are shown in Fig.\,\ref{fig_one_mode_tomo}c,d. 

Finally, we estimate the most likely state $\hat \rho$ of the mechanical resonator using convex optimization~\cite{Wang2019}. In this procedure, we perform Ramsey measurements (Fig.\,\ref{fig_one_mode_tomo}c)  to find $P_\alpha(n)$ for 36 different complex values of $\alpha$ (Fig.\,\ref{fig_one_mode_tomo}b). We then infer the most likely state $\hat{\rho}$ by minimizing the distance between the experimentally obtained $P_\alpha(n)$ and $\langle n |\hat D_\alpha \hat{\rho} \hat D^\dagger_\alpha|n\rangle$ over all measured $\alpha$. The reconstructed $\hat{\rho}$ are shown in Fig.\,\ref{fig_one_mode_tomo}e, and have state fidelities $\FidOne = \langle\psi|\hat{\rho}|\psi\rangle = 0.913{\,\pm\,}0.003$, $0.600{\,\pm\,}0.002$, and $0.811{\,\pm\,}0.002$ for the target phonon states $|\psi\rangle=\ket{0}$, $\ket{1}$, and $\ket{0}+\ket{1}$, respectively. These reconstructed $\hat{\rho}$ are then used to compute the Wigner functions $W(\alpha)$ in Fig.\,\ref{fig_one_mode_tomo}f, where the negative values in $W(\alpha)$ for $\ket{1}$ demonstrate the quantum nature of the phonon state. We note that the phonon parity can also simply be extracted from the measured $P_\alpha(n)$, from which we directly observe negative parity values in $\ket{1}$ (see methods and Fig.\,\ref{fig_SI_coarse_wigner}).

We attribute the imperfect overlaps $\FidOne$ between the target and measured states mostly to mechanical decay during the experiment. The Ramsey measurement duration is constrained by the time required to resolve the dispersive shifts, $\pi/\chi_2 \simeq$ 700\,ns, which is comparable to the decay $\Tonemtwo$. Quantum master equation simulations of the tomography protocol agree with the measured fidelities when we include the observed mechanical decoherences $\Tonemtwo$ and $\Ttwomtwo$ (see methods).

\begin{figure}[h!]
    \centering
    \includegraphics[width=89mm]{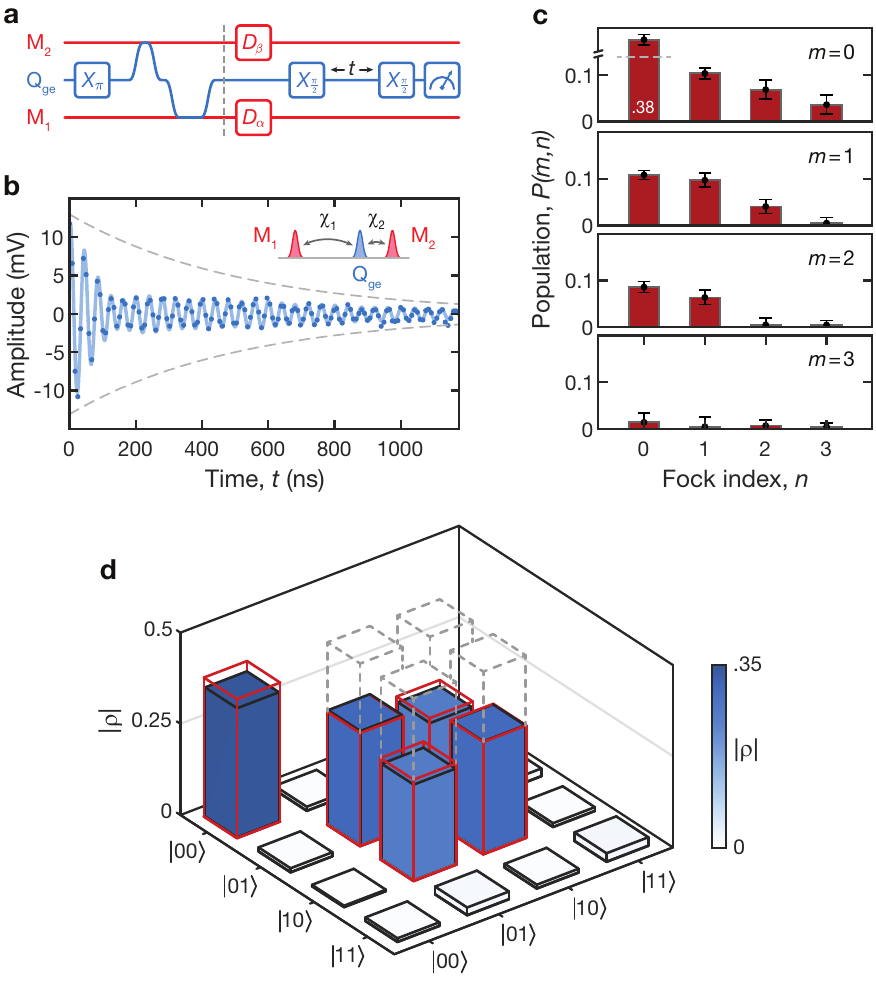}
    \caption{{\bf Joint tomography of a mechanical Bell-state.} {\bf a,}~Pulse sequence for mechanical Bell-state preparation (left) and two-mode tomography (right). The qubit is first excited to $\ket{e}$, followed by a $\sqrt{i\textsc{swap}}$ and $i\textsc{swap}$ operation to the mechanical resonators $M_2$ and $M_1$, respectively. Two-mode tomography is performed as in Fig.\,\ref{fig_one_mode_tomo}a, with the modification that a displacement $\hat{D}_\alpha\otimes\hat{D}_\beta$ is now applied simultaneously to each resonator.
    {\bf b,}~Representative Ramsey interference pattern (points) and fit (line) for the two-mode measurements. The qubit frequency is positioned between the mechanical resonators to achieve discernibly different dispersive shifts $\chi_1$ and $\chi_2$ from each mode (inset).
    {\bf c,}~Extracted joint phonon number distribution $P_{\alpha\beta}(m,n)$ from the fit of {\bf b}, with the $P_{\alpha\beta}(0,0) \simeq 0.38$ element truncated for visual clarity.
    {\bf d,}~Reconstructed density matrix $\hat{\rho}$ for the mechanical Bell-state. 25 different Ramsey measurements ({\bf a}-{\bf c}) are combined to obtain the most likely quantum state (blue bars), in good agreement with numerical simulations (red) that take into account mechanical decay during the Ramsey measurement of an initial ideal Bell-state (grey).}
    \label{fig_bell_state}
\end{figure}

By developing fast gates for multiple mechanical oscillators and extending our tomography protocol to bipartite states, we realize a small quantum acoustic processor that can generate and characterize entangled states of mechanical systems. As in Fig.\,\ref{fig_bell_state}a, our entangling gate consists of multiple sub-operations to create a mechanical Bell-state, $\ketBell = \ket{01} + e^{i\phi}\ket{10}$. After exciting the qubit, a $\sqrt{i\textsc{swap}}$ operation is performed between the qubit and upper mechanical mode to maximally entangle the two. The qubit state is then fully swapped to the lower mechanical mode, which translates the entanglement to be between the two mechanical systems. 

Next, we perform tomography on the joint mechanical system by extending our Ramsey measurement approach. We position the qubit frequency such that the mechanical dispersive shifts, $\chi_1/2\pi=(\text{-}517{\,\pm\,}6)\,\text{kHz}$ and $\chi_2/2\pi=(\text{-}799{\,\pm\,}5)\,\text{kHz}$, are distinguishable from each other, with a typical Ramsey measurement signal shown in Fig.\,\ref{fig_bell_state}b. In fitting these two-mode experiments' interference patterns, the model of Eq.\,\ref{eq_ramsey_fit} is extended to accommodate both resonators by replacing $A_n \rightarrow A_{mn}$ and $2\chi n \rightarrow 2\chi_1m + 2\chi_2n$ for Fock indices $m$ and $n$ of the lower and upper mechanics, respectively. Normalization of the signal amplitudes $A_{mn}$ gives the joint phonon number distribution $P_{\alpha\beta}(m,n)$ (Fig.\,\ref{fig_bell_state}c). In order to reconstruct the joint state $\hat{\rho}$ of the two mechanical systems, we repeat the experiment for 25 different combinations of displacements $\hat{D}_\alpha\otimes\hat{D}_\beta$ on the resonators, each time extracting the associated $P_{\alpha\beta}(m,n)$. We then estimate $\hat{\rho}$ from the set of $P_{\alpha\beta}(m,n)$ using convex optimization (see methods), resulting in the reconstructed state shown in Fig.\,\ref{fig_bell_state}d. The overlap of the inferred state with the target Bell-state is $\FidTwo = \braBell \hat{\rho} \ketBell = 0.57{\,\pm\,}0.02$, with a quantum state purity $\text{tr}(\hat{\rho}^2) = 0.46{\,\pm\,}0.02$. Numerical simulations of the mechanical system, shown in Fig.\,\ref{fig_bell_state}d, are in good agreement with the measured $\hat{\rho}$, allowing us to attribute the dominant source of loss in fidelity $\FidTwo$ to mechanical $\Tonem$ and $\Ttwom$ decay during the Ramsey measurement (see methods).

In conclusion, we demonstrate deterministic quantum control over a pair of nanomechanical resonators and characterize their joint quantum state using a dispersive, non-demolition measurement. In future work, mitigation of TLS-induced decoherence in lithium niobate phononic crystal resonators should allow for longer mechanical coherence times~\cite{Wollack2021,MacCabe2019}, which presently limit the observed state fidelities in our device. This experiment's flip-chip architecture is well-suited for separate optimization of the qubit and mechanical systems by enabling a modular approach to engineering hybrid quantum systems. Our hardware approach has enabled deterministic manipulation of quantum entanglement between macroscopic mechanical objects, and can be extended to architectures including quantum random access memories and biased-error cat qubits~\cite{Pechal2018,Hann2019,Chamberland2020}.

\section*{Methods}

\renewcommand{\theequation}{S\arabic{equation}}
\renewcommand{\thefigure}{S\arabic{figure}}
\renewcommand{\thetable}{S\arabic{table}}

\setcounter{equation}{0}
\setcounter{figure}{0}
\setcounter{table}{0}

\textbf{Fabrication.}
Our device fabrication closely follows previous methods~\cite{Arrangoiz-Arriola2019}, with the important difference that the processes for qubits and nanomechanical structures are now performed on separate dies. We have moved to a slightly different material platform for the mechanics chip, in which the thin-film LN has been doped with magnesium oxide (MgO) to improve the mechanical properties of the crystal~\cite{Wollack2021}. Additionally, we thermally anneal the mechanics chip (8 hours at 500 C) before patterning the device. On the qubit chip, we have added aluminum crossovers across the qubit control lines and readout transmission line. We have also developed an oxygen plasma descum process to remove polymer residues from inter-metallic layers, reducing TLS-induced microwave loss. 

\begin{figure*}[!htb]
    \centering
    \includegraphics[width=183mm]{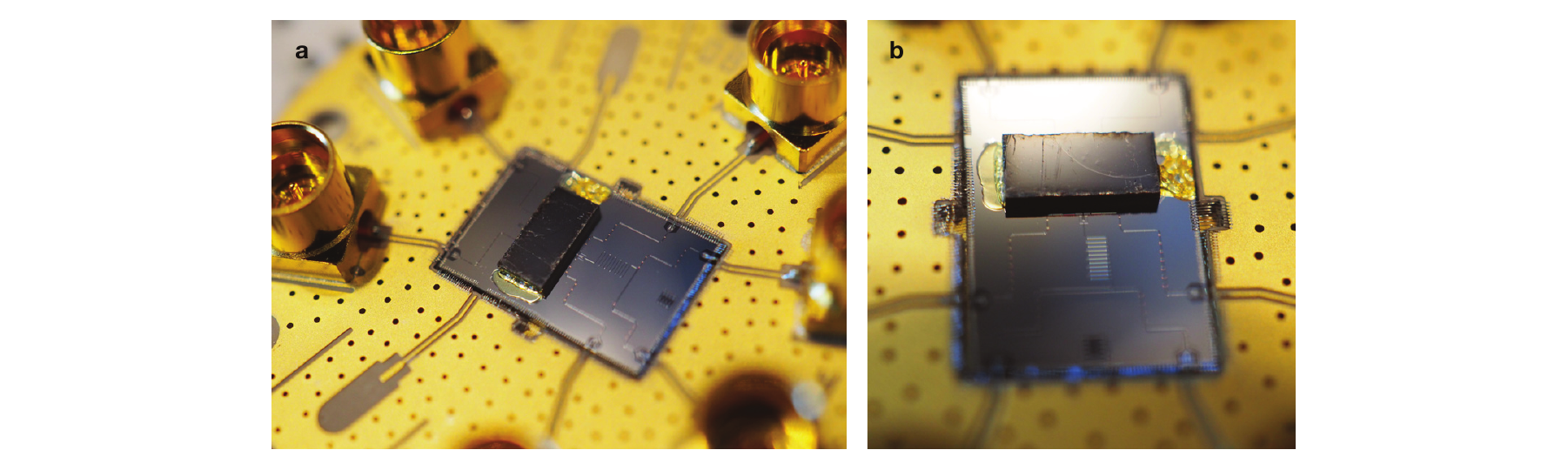}
    \caption{{\bf Device images.} {\bf a,} Angled top view and {\bf b,} angled side view photographs of the fully packaged device. The top (mechanics) chip is secured face-down to the bottom (qubit) chip by an adhesive polymer (9:1 ethanol to GE varnish) applied manually to the sides of the chip. }
    \label{fig_SI_chip_pics}
\end{figure*}

The flip-chip bonding procedure is the final step in our fabrication process. We use a submicron die bonder (Finetech) to align the two chips by positioning the two pads of the coupling capacitor on top of each other. The two chips’ active surfaces are brought to a separation distance of 1 $\mu$m, as allowed by 500 nm aluminum spacer ridges patterned on each chip. Finally, an adhesive polymer (9:1 ethanol/GE varnish) is manually applied to the outer edges of the top chip to secure it in place. Images of the final integrated flip-chip device are shown in Fig.\,\ref{fig_SI_chip_pics}.

\textbf{Mechanics design.}
For our experiment, it is important to carefully choose the frequency arrangement such that all modes are sufficiently protected from decoherence channels, while the qubit-mechanics interaction remains in the dispersive regime. Using finite-element simulations, we choose a phononic crystal geometry with a bandgap extending from approximately 1.90 to 2.50 GHz and a pitch of $a = 900$\,nm. The mechanical frequencies, controlled by adjusting the width of the defect site, are designed to be approximately 150 MHz away from the bandgap edges, ensuring that the modes are protected from clamping losses.

\begin{figure*}[!htb]
    \centering
    \includegraphics[width=183mm]{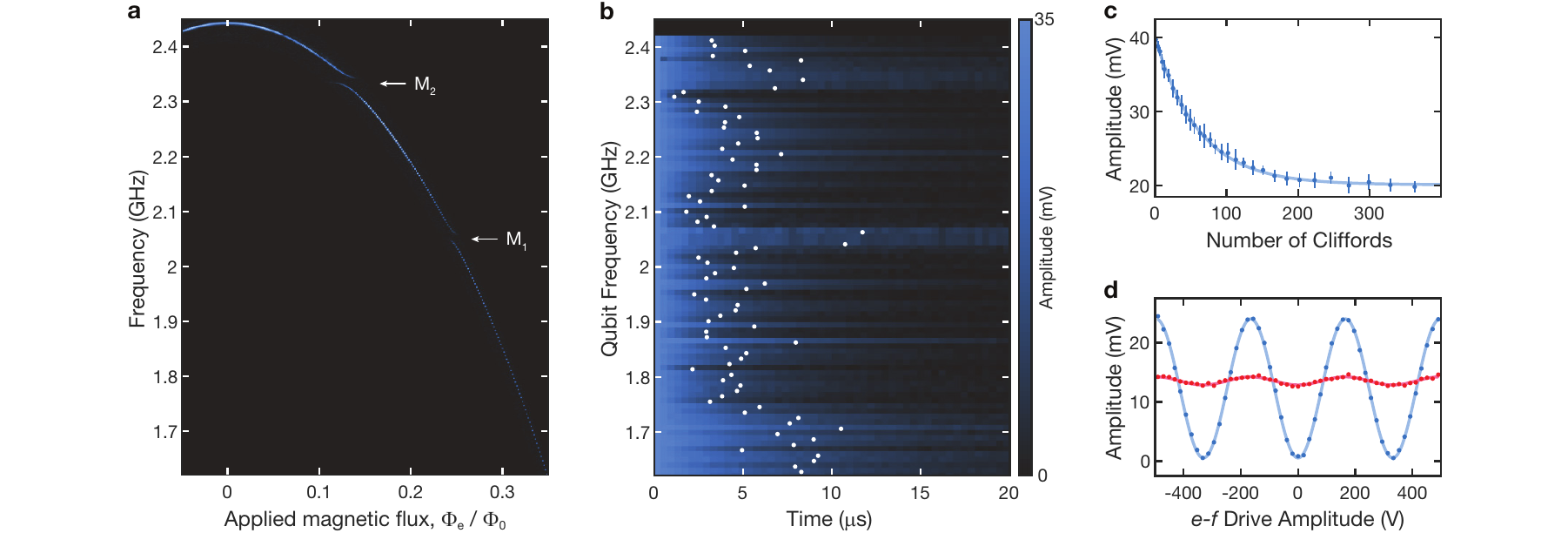}
    \caption{{\bf Qubit characterization.} {\bf a,} Qubit spectrum as a function of the externally applied magnetic flux $\Phi_e$, in units of the magnetic flux quantum $\Phi_0$. The qubit is tuned from its maximum frequency $\omegage^\text{max}/2\pi = 2.443\,$GHz and shows avoided crossings at $\omegamone/2\pi = 2.053\,$GHz and $\omegamtwo/2\pi = 2.339$\,GHz, corresponding to the mechanical modes $M_1$ and $M_2$.
    {\bf b,} Qubit $T_1$ as a function of qubit frequency. Each horizontal slice represents a separate ringdown measurement, with qubit excited state probability indicated by the color bar. The fitted $T_1$ values for each slice are plotted as white points.
    {\bf c,}~Randomized benchmarking results. Here, the qubit response is measured after applying a random sequence of Clifford gates, with the final Clifford always chosen to map the sequence's the cumulative effect to $\ket{e}$.
    {\bf d,}~Qubit thermometry measurement. The dark blue points (dark red points) show the measurement result with (without) an initial $X_\pi$ pulse to exchange the steady-state $\ket{g}$ and $\ket{e}$ populations. The light blue (light red) lines show the fits for these Rabi-like oscillations, with amplitudes $A_g$ ($A_e$). From these amplitudes, we estimate a qubit thermal population in $\ket{e}$ of $P_{e,\text{th}} = 0.057.$}
    \label{fig_SI_qubit_characterization}
\end{figure*}

\textbf{Qubit design and control.}
Our device utilizes a transmon-style qubit with microwave control lines and a dispersively coupled microwave resonator for readout. An on-chip flux line positioned near the qubit's SQUID loop provides capability for both static (DC) and rapid (pulsed) frequency tuning of the qubit via externally applied magnetic flux. In Fig.\,\ref{fig_SI_qubit_characterization}a, we measure the qubit's frequency tuning curve, with the maximum qubit frequency at $\omegage^{\text{max}}/2\pi = 2.443$\,GHz. Our device has charging energy $E_C/h = \alpha_q/2\pi$ = 126\,MHz and Josephson energy $E_J/h$ = 6.550\,GHz, ensuring that it operates well into the transmon regime~\cite{Koch2007}, $E_J/E_C \gg 1$.

For tomography experiments, the qubit operating frequency is chosen to be in between the two mechanical frequencies. This ensures that both the primary qubit transition $\omegage$ and the next higher transition $\omega_{\text{ef}} = \omegage - \alpha_q$ are sufficiently distant from the mechanical modes that the qubit is effectively decoupled, allowing us to perform rotations of the qubit state with high fidelity. Placing the qubit frequency in this region also gives strong dispersive coupling to both mechanical modes in order to perform joint tomography of the mechanical systems. 

We perform gates on the qubit state by applying microwave pulses with variable amplitude, phase, and duration to the qubit's $XY$ line. For these experiments, we use 20\,ns DRAG pulses with approximately Gaussian envelopes~\cite{Motzoi2009,Chen2016}. Utilizing randomized benchmarking techniques~\cite{Magesan2011,Corcoles2013}, we observe a single qubit gate fidelity of 0.996 at the operating frequency used for tomography ($\omegage/2\pi = 2.26\,\text{GHz}$) as shown in Fig.\,\ref{fig_SI_qubit_characterization}c. To measure the qubit state, we use a standard cQED approach of dispersive readout through an off-resonantly coupled coplanar waveguide resonator. To infer the qubit excited state probability, we apply a microwave pulse to the readout resonator's transmission line and measure the scattered response, which allows us to detect shifts in its resonant frequency induced by the qubit's state.

In Fig.\,\ref{fig_SI_qubit_characterization}b, we measure the qubit energy decay time $T_1$ over a large frequency range and plot the results. For each horizontal slice, we statically bias the qubit to the indicated frequency and perform a standard ringdown measurement to study its $T_1$ energy decay. The qubit excited state probability, indicated by the color bar, is plotted as a function of time. The white points show the resulting $T_1$ values, extracted by fitting each data slice to an exponential decay function. From this data set, we find the average qubit decay time to be $T_{1,\text{avg}} = (4.9 \pm 2.3)\,\mu$s.

We also characterize the thermal population of the qubit with a thermometry experiment, shown in Fig.\,\ref{fig_SI_qubit_characterization}d. For this measurement, we use a Rabi population method~\cite{Satzinger2018, Geerlings2013} to quantify the residual qubit population in $\ket{e}$. This is done by driving rotations of the qubit state between the $\ket{e}$ and $\ket{f}$ levels with varying rotation angle. A final $X_\pi$ pulse exchanges the $\ket{g}$ and $\ket{e}$ populations before we measure the qubit state. We perform this measurement both with and without an optional $X_\pi$ pulse at the beginning of the sequence, which exchanges the steady-state $\ket{g}$ and $\ket{e}$ populations in the qubit. These measurements produce two Rabi-like oscillation patterns whose amplitudes $A_g$ and $A_e$ contain information about the thermal $\ket{e}$ population, $P_{e,\text{th}} = A_e/(A_g + A_e)$. By this method, we find $P_{e,\text{th}} = 0.057$ with the qubit biased to $\omegage/2\pi = 1.798$\,GHz. We perform this measurement far detuned from both mechanical modes to ensure high fidelity rotations of the qubit states. At this operating point, the reported value is likely an upper bound on the qubit thermal population relevant for our primary experiments.

\textbf{Master equation simulations.}
To model the quantum dynamics of our device and obtain estimates of the mechanical state fidelities, we perform time-domain master equation simulations using the QuTiP package~\cite{qutip}. First, we simulate the qubit-mechanics dynamics during Rabi-swap experiments to illustrate how nanosecond-timescale flux pulses can be used to manipulate the device. Here, the qubit is modeled as a 3-level nonlinear resonator $\hat{a}$ with time-dependent frequency $\omegage(t)$ and anharmonicity $\alpha_q$, subject to $T_1$ decay and pure dephasing $T_\phi$. The mechanical resonators are taken to be 3-level harmonic oscillators $\hat{b}_i$, also with $\Tonem$ and $T_{\phi,\text m}$ decoherence channels. For this experiment, the total Hamiltonian $\hat{H} = \hat{H}_0 + \hat{H}_\text{int} + \hat{H}_\text{d}$ has contributions
\begin{align*}
    &\hat{H}_0 = \omegage(t)\hat{a}^\dagger\hat{a} - \frac{\alpha_q}{2}\hat{a}^\dagger\hat{a}^\dagger\hat{a}\hat{a} + \sum_i\omegami\hat{b}_i^\dagger\hat{b}_i\,,
    \\ &\hat{H}_\text{int} = \sum_i g_i(\hat{a} + \hat{a}^\dagger)(\hat{b}_i + \hat{b}_i^\dagger)\,,
    \\ &\hat{H}_\text{d} = \tfrac{1}{2}\left(\Omega(t)\hat{a} + \Omega^*(t)\hat{a}^\dagger \right) \,.
\end{align*}
A drive term $\hat{H}_\text{d}$ allows for population of either the qubit or the mechanical resonators, depending on the chosen modulation frequency of the applied drive, $\Omega(t)$. We also allow applied flux pulses to add time-dependent frequency control of the qubit, $\omegage(t)=\omegage + \tilde{\omega}_\text{ge}(t)$, where $\omegage$ is the static qubit frequency and $\tilde{\omega}_\text{ge}(t)$ represents the transient frequency control.

In simulating the time evolution of the total quantum system $\hat{\rho}$, we numerically integrate the Lindblad master equation, 
\begin{align*}
    \frac{d\hat{\rho}}{dt} = -i[\hat{H},\hat{\rho}] + \sum_k\left(\hat{c}_k\hat{\rho}c_k^\dagger - \frac{1}{2}\{\hat{c}_k^\dagger \hat{c}_k, \hat{\rho}\}\right)\,,
\end{align*}
with collapse operators $\hat{c}_k = \hat{b}_1/\sqrt{\Tonemone}$, $\hat{b}_2/\sqrt{\Tonemtwo}$, $\hat{b}_1^\dagger \hat{b}_1/\sqrt{2T_{\phi,\text{m}_1}}$, $\hat{b}_2^\dagger \hat{b}_2/\sqrt{2T_{\phi,\text{m}_2}}$, $\hat{a}/\sqrt{T_1}$, and $\hat{a}^\dagger\hat{a}/\sqrt{2T_\phi}$. Note that all model parameters are experimentally determined from standard qubit measurements, with a subtlety in our choice of the bare frequencies $\omegage$ and $\omegami$. In order to obtain consistent results between experiment and theory, $\hat{H}$ is first defined in the bare basis, then diagonalized to find the dressed basis eigenstates and eigenvalues. The bare frequencies of $\hat{H}$ are then chosen such that the dressed frequencies of the qubit and mechanics best match the experimentally measured values at steady state. These dressed states can then be used to evaluate final state probabilities and expectation values. 

\begin{figure*}[!htb]
    \centering
    \includegraphics[width=183mm]{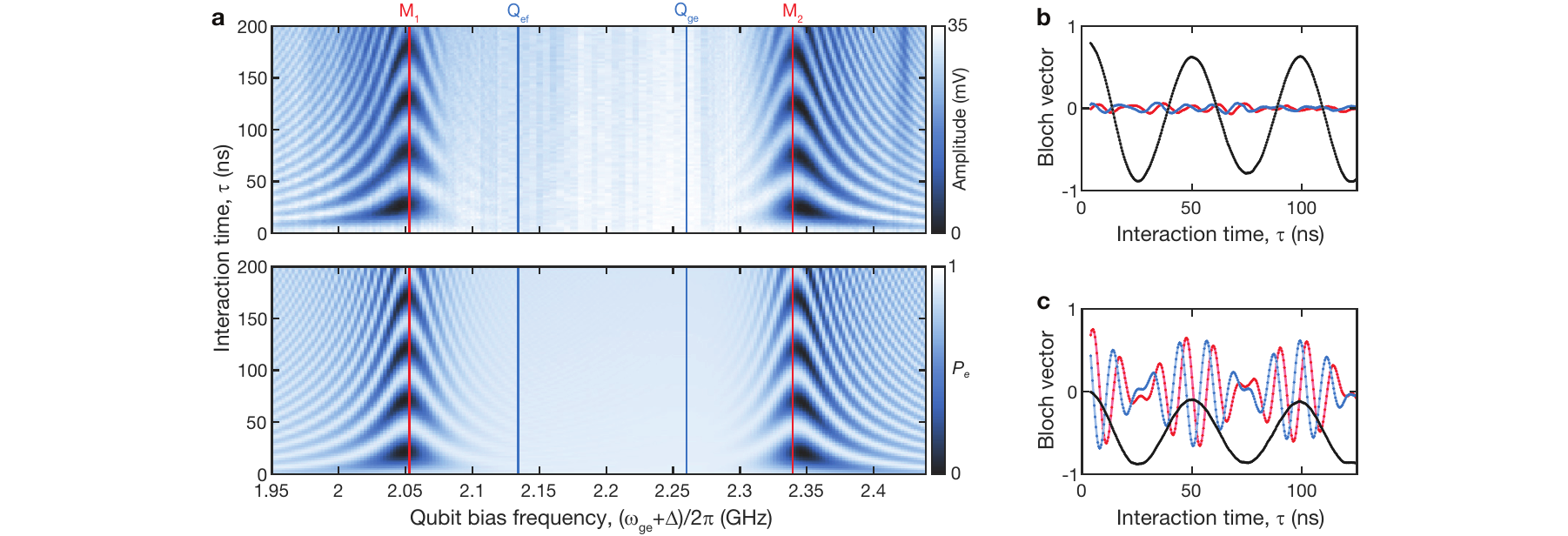}
    \caption{{\bf Swap characterization.} {\bf a,}~Experimental results (top) and simulated qubit excited state probability $P_e$ (bottom) for the Rabi-swap experiment described in Fig.\,\ref{fig_characterization}b,d of the main text. The asymmetry in the chevrons of the qubit-mechanics interaction is replicated by master equation simulations.
    {\bf b,}~Qubit state tomography results for the $X$ (blue), $Y$ (red), and $Z$ (black) components of the qubit state Bloch vector during a resonant Rabi-swap experiment. The qubit is initially prepared in $\ket{e}$, then swapped to the upper mechanical mode $M_2$ before performing tomography on the qubit.
    {\bf c,}~Qubit state tomography results for a resonant Rabi-swap experiment, similar to {\bf b}, where the qubit now starts in $\ket{g}+\ket{e}$.}
    \label{fig_SI_swap_characterization}
\end{figure*}

Using this framework, we aim to reproduce the asymmetry present in the qubit-mechanics chevrons of Fig.\,\ref{fig_characterization}d in the main text. These simulations show excellent agreement with experimental Rabi-swap results, as shown in Fig.\,\ref{fig_SI_swap_characterization}a. The limited visibility of the fringes near the static qubit bias $\omegage$ is due to the details of the dressed state in a system where the coupling $g_i$ is always present. In the case of small detuning $\tilde{\omega}_\text{ge}(t)$, the instantaneous dressed bases do not change appreciably, and so the qubit-like mode roughly remains in the same dressed eigenstate throughout the operation. This is in contrast to systems where the $g_i$ can be turned off during single qubit operations, thereby avoiding dressing of the qubit state except when the coupling is desired.

We perform additional master equation simulations to compare with the observed mechanical state fidelities $\FidOne$ and $\FidTwo$ reported in the main text. For these simulations, we assume the initial mechanical state $\hat{\rho}(0)$ is the ideal target state, then let $\hat{\rho}(t)$ freely evolve during the Ramsey measurement, subject only to mechanical collapse operators $\hat{c}_k = \hat{b}_i/\sqrt{T_{1,m_i}}$ and $\hat{b}_i^\dagger \hat{b}_i/\sqrt{2T_{\phi,m_i}}$. Note that we now ignore any dynamics of the qubit, as well as the details of the state preparation. For the single-mode states $\ket{1}$ and $\ket{0}+\ket{1}$, we obtain fidelities 0.566 and 0.809 from simulation, similar to the measured $\FidOne$ of $0.600{\,\pm\,}0.002$ and $0.811{\,\pm\,}0.002$. For the mechanical Bell-state, master equation simulations give a fidelity of 0.587, compared to the observed $\FidTwo = 0.57{\,\pm\,}0.02$. The good agreement between simulation and our measured fidelities suggest that $\FidOne$ and $\FidTwo$ are likely limited by $\Tonem$ and $\Ttwom$ decoherence mechanisms in the mechanical system.

\textbf{Swap characterization.}
To demonstrate control of the qubit-mechanics swap operation, we perform quantum state tomography on the qubit during resonant Rabi-swap experiments. Using the pulse sequence of Fig.\,\ref{fig_characterization}b, we bring the qubit into resonance with the upper mechanical mode $M_2$ for an interaction time $\tau$ before applying tomography gates $X_\theta$ or $Y_\theta$, chosen from $\theta = \{0, \pm\pi/2,\pm \pi\}$. Combining the results of this set of measurements allows for the reconstruction of the qubit Bloch vector $\langle \vec{\sigma} \rangle$, shown in Fig.\,\ref{fig_SI_swap_characterization}b,c. Note that since our experiment does not have single-shot qubit state readout, we calibrate the observed qubit response using the measured qubit thermal population $P_{e,\text{th}}$ in order to estimate the Bloch vector. In Fig.\,\ref{fig_SI_swap_characterization}b, the qubit is prepared in $\ket{e}$ before resonantly interacting with $M_2$; the resulting data show the excitation periodically returning to the qubit, with the $\sigma_X$ and $\sigma_Y$ components largely unaffected. A similar experiment is performed in Fig.\,\ref{fig_SI_swap_characterization}c, with the qubit now starting in the superposition $\ket{g} + \ket{e}$. As expected, the qubit's superposition is recovered from the mechanical resonator at even multiples of the swap time.

\textbf{Time domain data analysis.}
The Ramsey measurements used for state tomography contain information about the phonon number distribution of the dispersively coupled mechanical state. The distinct spectral components in the number-split qubit spectrum create an interference pattern which depends strongly on the mechanical occupation, as shown in Fig.\,\ref{fig_one_mode_tomo}b. We fit the data to a function of the form (Eq.\,\ref{eq_ramsey_fit} in main text)
\begin{align*}
    S(t) = \sum_{n=0} A_n e^{-\kappa t/2} \cos [(\omega_0 + 2 \chi n) t + \varphi_n]\,,
\end{align*}
where $\chi$, $\kappa$, and $A_n$ are model fit parameters. In $S(t)$, the component corresponding to the $n$th Fock level's occupation is given an amplitude $A_n$ and frequency $\omega_0 + 2 \chi n$. Here, the dispersive shift $\chi$ is constrained to be the same for all $n$, and the frequency $\omega_0/2\pi = 25\,\text{MHz}$ is the programmed frame detuning of the Ramsey sequence's second $\Xpihalf$ pulse. The phases $\varphi_n = 2\chi n (2\tilde{\tau})$ account for qubit phase accumulation during the two $\Xpihalf$ pulses of the Ramsey sequence, each with a pulse duration $\tilde{\tau} = 20\,\text{ns}$. Since we define $S(t)$ in terms of the elapsed time $t$ between the Ramsey pulses, there is a small phonon-state dependent phase accumulation at $t=0$ due to the finite operation time of our single qubit gates. The exponential decay term $e^{-\kappa t/2}$ represents an effective dephasing time which is dominated by the qubit $T_2$, but also includes a contribution from the mechanical state. This fit allows us to extract the Fock populations $P_\alpha(n)$ by normalizing the spectral amplitudes: $P_\alpha(n) = A_n / \Sigma_n A_n$.

To extend the model to the two-mode case, we replace $A_n \rightarrow A_{mn}$ and $2\chi n \rightarrow 2\chi_1m + 2\chi_2n$ for Fock indices $m$ and $n$ of the resonators. We also need to adjust the zero-delay qubit phase $\varphi_n \rightarrow \varphi_{mn} = (2\chi_1m + 2\chi_2n) (2\tilde{\tau})$ to account for both resonators shifting the qubit frame during the Ramsey $\Xpihalf$ pulses. This yields an adjusted time domain model
\begin{align*}
    S(t) = \sum_{m,n} A_{mn} e^{-\kappa t/2} \cos[(\omega_0\!+\!2\chi_1m\!+\!2\chi_2n) t + \varphi_{mn}].
\end{align*}
We use this function to fit the Ramsey measurements of Fig.\,\ref{fig_bell_state}b in the main text, and thereby determine the two-mode phonon number distribution $P_{\alpha\beta}(m,n) = A_{mn} / \Sigma_{m,n} A_{mn}$.

\textbf{State reconstruction.}
For both the single- and two-mode tomography demonstrated in this experiment, we can use similar protocols for state reconstruction. We perform tomography of an unknown two-mode mechanical state $\hat{\rho}$ by applying displacements $\hat{D}_{\alpha\beta} \equiv \hat{D}_\alpha \otimes \hat{D}_\beta$ on $\hat{\rho}$, then measuring the diagonal elements $P_{\alpha\beta}(m,n)$ of the resulting joint state $\hat{\rho}(\alpha,\beta) = \hat{D}_{\alpha\beta}{\,\hat{\rho}\,} \hat{D}^\dagger_{\alpha\beta}$~\cite{Wang2019}. More specifically, the set of measurement data can be represented as $P_{\alpha\beta} (m,n) = \langle m,n | \hat{D}_{\alpha\beta}{\,\hat{\rho}\,} \hat{D}^\dagger_{\alpha\beta} | m,n \rangle$, where $|m,n\rangle = |m\rangle \otimes |n\rangle$ is the joint Fock basis of the resonators. In our experiment, we choose from combinations of complex displacements with amplitudes $(|\alpha|,|\beta|)\le (0.6,0.7)$, and fit the resulting $P_{\alpha\beta} (m,n)$ up to maximum Fock indices $(m_\text{max},n_\text{max})= (3,3)$. The unknown density matrix $\hat{\rho}$ can then be reconstructed by minimizing the loss function
\begin{align*}
\mathcal{L} (\hat{\rho}) = \sum_{\alpha,\beta} \sum_{m,n} \left| \langle m,n | \hat{D}_{\alpha\beta}{\,\hat{\rho}\,} \hat{D}^\dagger_{\alpha\beta} | m,n \rangle - P_{\alpha\beta} (m,n) \right|^2 ,
\end{align*}
a convex problem that can be solved efficiently using the CVX package~\cite{cvx}. The single-mode mechanical state tomography follows a similar method by setting either $\alpha$ or $\beta$ to zero and ignoring the corresponding mode in the analysis. For the single-mode tomography experiments, we use $|\alpha|\le 1.25$ and $n_\text{max} = 8$.

\textbf{Direct parity calculation.}
In our single-mode tomography protocol, we choose to reconstruct the Wigner functions by fitting over candidate $\hat{\rho}$ for experimental efficiency. However, the mechanical resonator's Wigner function $W(\alpha)$ can also be directly computed from the parity $\Pi(\alpha) = \Sigma_{n} (-1)^n P_\alpha(n) = \tfrac{\pi}{2}W(\alpha)$ of each Ramsey measurement. In a separate data set  (Fig.\,\ref{fig_SI_coarse_wigner}), we measure $\Pi(0)=-0.36$ for the $|\psi\rangle = |1\rangle$ target state, which confirms the quantum nature of this prepared state by direct measurement.

\begin{figure*}[!htb]
    \centering
    \includegraphics[width=183mm]{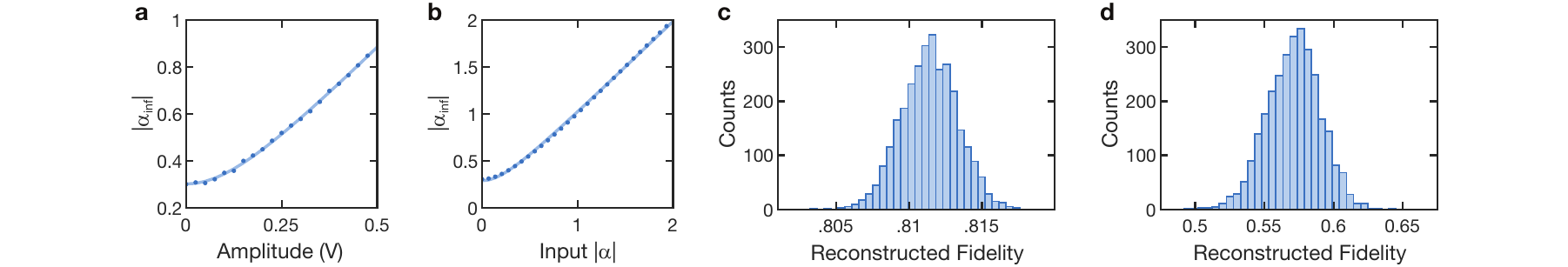}
    \caption{{\bf State reconstruction.} {\bf a,}~Experimental results for the mechanical displacement  calibration, showing the inferred  displacement amplitudes $|\alpha_\text{inf}|$ (points) corresponding to each applied pulse's voltage amplitude, and a fit to the hockey-stick model (line).
    {\bf b,}~Simulation and fit of the displacement calibration performed experimentally in {\bf a}. A small thermal state ($P_\text{th} = 0.10$) is displaced with a programmed amplitude (input $|\alpha|$, x-axis), from which we determine the inferred displacement amplitude ($|\alpha_\text{inf}|$, y-axis). The simulation data are plotted in dark blue points, with a fit to the hockey-stick model plotted in light blue. In the linear portion of the graph, we find the ratio of these values, $c_1=$ (inferred $|\alpha|$)/(input $|\alpha|$) $\simeq 1$, as expected.
    {\bf c,}~Results of error propagation for the reconstructed fidelity $\FidOne = 0.811{\,\pm\,}0.002$ in single-mode tomography of the $\ket{0} + \ket{1}$ state. {\bf d,}~Results of error propagation for the joint tomography Bell-state fidelity $\FidTwo = 0.57{\,\pm\,}0.02$.}
    \label{fig_SI_state_reconstruction}
\end{figure*}

\begin{figure*}[!htb]
    \centering
    \includegraphics[width=183mm]{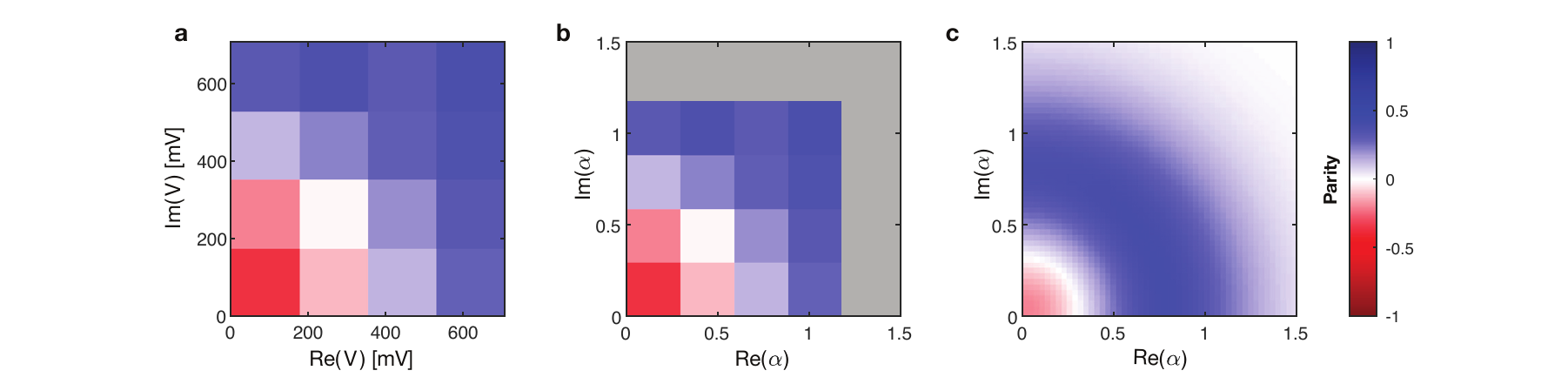}
    \caption{{\bf Direct parity measurement.} {\bf a,} Parity of the displaced mechanical state $\ket{1}$ prepared in the upper mechanical mode. The state is prepared and characterized using the same pulse sequence as in Fig.\,\ref{fig_one_mode_tomo}a. Here, we use 16 displacements in the upper-right quadrant of the complex plane and perform the Ramsey measurement for a total time $t_\text{max} = 4.0\,\mu$s. For each displacement $\hat{D}(\alpha)$, parity is computed from the extracted phonon number distribution as $\Pi(\alpha) = \Sigma_{n=0} (-1)^n P_\alpha(n)$. The axes correspond to the complex voltage amplitudes of the applied microwave pulses that generate these displacements. This yields a minimum observed parity $\Pi(0) \simeq -0.36$ at the origin.
    {\bf b,} The same parity measurement, plotted in terms of the inferred displacement amplitudes. We extrapolate these $\alpha$ values using the calibration scheme described in Fig.\,\ref{fig_SI_state_reconstruction}.
    {\bf c,} Upper-right quadrant of the reconstructed Wigner function shown in Fig.\,\ref{fig_one_mode_tomo}f, reproduced here for comparison. }
    \label{fig_SI_coarse_wigner}
\end{figure*}

\textbf{Displacement calibration.}
Mechanical state reconstruction relies on knowing the amplitudes $|\alpha|$ of the displacements $\hat{D}_\alpha$ that we apply during each tomography pulse sequence. This requires a calibration relating the applied microwave pulse's voltage amplitude to the resulting mechanical displacement amplitude $|\alpha|$, shown in Fig.\,\ref{fig_SI_state_reconstruction}a. Here, we displace the upper mechanical mode by applying a microwave pulse to the qubit $XY$ line at the mechanical frequency, then perform Ramsey interferometry on the resulting state to extract the phonon number distribution $P_\alpha (n)$. Next, we perform a least squares fit to find the coherent state $\ket{\alpha}$ whose coefficients $|\langle n|\alpha\rangle|^2$ most closely match the measured $P_\alpha (n)$ to obtain the inferred displacement amplitude $|\alpha_\text{inf}|$. We find that the relation between the applied voltage amplitude $V$ and the inferred displacement amplitude $|\alpha_\text{inf}|$ follows a ``hockey-stick" curve $|\alpha_\text{inf}| = ((c_1V)^2+c_2)^{1/2}$, where the second fit parameter $c_2$ accounts for the thermal population of the mechanical mode. During state reconstruction, we only attribute the displacement amplitude to the voltage we apply, namely $|\alpha| = c_1 V$. For our system, we find $c_1 = 1.664 \pm 0.005$ and $c_2 = 0.091 \pm 0.001$.

We also perform a simulation of this displacement calibration procedure, and the results (Fig.\,\ref{fig_SI_state_reconstruction}b) show good agreement with the experimental data. In these simulations, a small thermal state $\hat{\rho}_\text{th}$ with population $P_{\text{th}} = 0.10$ is coherently displaced $\hat{D}_\alpha \hat{\rho}_\text{th} \hat{D}_\alpha^\dagger$ with a programmed amplitude $|\alpha|$. The phonon number distribution of the resulting state is fit to the nearest coherent state, from which we infer the effective displacement amplitude $|\alpha_\text{inf}|$. This operation yields a similar hockey-stick behavior as the experimental data, with a y-intercept $|\alpha_\text{inf}|= 0.305$. This corresponds to an average phonon number $n_\text{avg} = 0.093 \simeq P_\text{th}$. Fitting this simulated data to the hockey-stick model yields a scale factor between the inferred $|\alpha|$ and the input $|\alpha|$ of $c_1 = 0.984 \pm 0.002 \simeq 1$, as we expect.

\textbf{Error analysis of state reconstruction.}
We use Monte Carlo error propagation to determine the robustness of the mechanical state reconstruction previously described. In fitting the $P_\alpha(n)$ or $P_{\alpha\beta}(m,n)$ from $S(t)$, we obtain an estimate for the model parameters' covariance matrix during the nonlinear least squares regression. For error propagation testing, we then randomly resample the $P_\alpha(n)$ or $P_{\alpha\beta}(m,n)$ and displacement calibrations, using their respective statistical uncertainties computed from the model's covariance matrix. The resampled parameters are then fed into the convex optimization routine that minimizes $\mathcal{L}(\hat{\rho})$ to obtain the reconstructed state $\hat{\rho}$. This resampling process is repeated $3\times10^3$ times to obtain the resulting variations in the reconstructed density matrix fidelities shown in Fig.\,\ref{fig_SI_state_reconstruction}c,d. We use the standard deviation of these reconstructed fidelities to obtain the error estimates for the state fidelities $\FidOne$ and $\FidTwo$ reported in the main text.

\section*{Acknowledgments}
The authors would like to thank M. Kang, T.P. McKenna, W. Jiang, Y.P.Zhong, K.K.S. Multani, N.R. Lee, M.M. Fejer, and P.J. Stas for discussions. We acknowledge the support of the David and Lucille Packard, and Sloan Fellowships. This work was funded by the U.S. government through the Office of Naval Research (ONR) under grant No.~N00014-20-1-2422, the U.S. Department of Energy through Grant No. DE-SC0019174, and the National Science Foundation CAREER award No.~ECCS-1941826. E.A.W. was supported by the Department of Defense through the National Defense \& Engineering Graduate Fellowship. A.Y.C. was supported by the Army Research Office through the Quantum Computing Graduate Research Fellowship as well as the Stanford Graduate Fellowship. Device fabrication was performed at the Stanford Nano Shared Facilities (SNSF), supported by the National Science Foundation under award ECCS-2026822, and the Stanford Nanofabrication Facility (SNF). The authors wish to thank NTT Research for their financial and technical support.

\section*{Author Contributions}
E.A.W. and A.Y.C. designed and fabricated the device. E.A.W., A.Y.C., R.G.G., and P.A.A. developed the fabrication process. Z.W., R.G.G., and A.H.S.-N. provided experimental and theoretical support. E.A.W. and A.Y.C. performed the experiments and analyzed the data. E.A.W., A.Y.C. and A.H.S.-N. wrote the manuscript, with all others assisting. A.H.S.-N. supervised all efforts.

\section*{Additional Information}
P. Arrangoiz-Arriola is currently a research scientist at Amazon, and A. H. Safavi-Naeini is an Amazon Scholar. The other authors declare no competing financial interests. Correspondence and requests for materials should be addressed to A. H. Safavi-Naeini (safavi@stanford.edu)

\bibliography{LINQS_CQAD_Papers.bib}

\pagebreak
\clearpage

\begin{table}
    \centering
    \begin{ruledtabular}
    \begin{tabular}{@{\hspace{1cm}} c @{\hspace{0cm}} c @{\hspace{1cm}}}
     Parameter & Value(s) \\[0.5 ex]
    \hline
        $\omegage^\text{max}/2\pi$ & 2.443 GHz \Tstrut \\
        $\alpha_q/2\pi$ & 126 MHz  \\
        $T_1$ & $(4.9{\,\pm\,}2.3)\,\mu\text{s}$ \\
        $T_2$ (flux sweet spot) & 1.4 $\mu$s \\
        $T_2$ ($\omegage/2\pi = 2.26$\,GHz) & 0.8 - 1.2 $\mu$s \\
        $\omegami/2\pi$ & 2.053, 2.339 GHz \\
        $\Tonem$ & 1.23, 0.99 $\mu$s \\
        $\Ttwom$ & 0.87, 1.71 $\mu$s \\
        $g_i/2\pi$ & 9.5, 10.5 MHz \\
        $\omega_\text{r}^\text{max}/2\pi$ & 2.872 GHz \\
        $\kappa_r/2\pi$ & 1.29 MHz \Bstrut \\
    \end{tabular}
    \end{ruledtabular}
    \caption{{\bf Device parameters.} Parameters of the qubit, mechanical modes, and readout resonator. }
    \label{tab:dev_params}
\end{table}

\end{document}